\documentclass[10pt,conference]{IEEEtran}

\usepackage[utf8]{inputenc}
\usepackage{float}
\usepackage{booktabs} 
\usepackage{amsmath}
\usepackage{amssymb}
\usepackage{listings} 
\usepackage{graphicx}
\usepackage{nameref}
\usepackage{multirow}
\usepackage[inline]{enumitem}
\usepackage{array}
\usepackage{multicol}
\usepackage{graphicx}
\usepackage{diagbox}
\usepackage[table,x11names]{xcolor}
\usepackage{listings}
\usepackage{wrapfig}
\usepackage{lscape}
\usepackage{rotating}
\usepackage{epstopdf}
\usepackage{mathtools}
\usepackage{makecell}
\usepackage{tikz}
\usepackage[framemethod=TikZ]{mdframed}
\usepackage[bottom]{footmisc}
\usepackage{caption}
\usepackage{subcaption}
\usepackage[flushleft]{threeparttable}
\usepackage[numbers,sort&compress]{natbib} 
\usepackage{hyperref}
\usepackage{cleveref}

\usepackage[zerostyle=b,scaled=.85]{newtxtt}

\hypersetup{
    colorlinks=true,
    linkcolor=blue,
    urlcolor=blue,
    citecolor=blue,
    linktocpage=false
}

\lstdefinestyle{scala}
{%
    emph=[1]%
    {%
        groupBy,
        count,
        map,
        union,
        mapValues
    },
    emphstyle=[1]{\color{blue}},
    emph=[2]
    {%
        String,
        Path,
        Unit,
        List,
        Iterable,
        Array,
        Set,
        Seq,
        RDD,
        Map,
        Int,
        Long
    },
    emphstyle=[2]{\color{violet}},
    emph=[3]%
    {%
        ParsedLine,
        Dataset,
        MultiDataset
    },
    emphstyle=[3]{\color{teal}}
}

\lstdefinestyle{scala-types-only}
{%
    emph=[2]
    {%
        String,
        Iterable,
        Array,
        Set,
        RDD,
        Map,
        Int,
        Long
    },
    emphstyle=[2]{\color{violet}},
    emph=[3]%
    {%
        ParsedLine,
        Dataset,
        MultiDataset
    },
    emphstyle=[3]{\color{teal}}
}

\newcommand{\mynumold}[1]{{\oldstylenums{#1}}}

\newcounter{lstNoteCounter}

\newcommand*\lnnum[1]{\tikz[baseline=(char.base)]{
            \node[shape=circle,draw,inner sep=0.8pt,
                        fill=black, text=white] (char) { \rmfamily\bfseries\scriptsize#1};}}

\lstnewenvironment{annotatedcsource}[3][scala]
{%
    \setcounter{lstNoteCounter}{0}
    \lstset{
        style=#1,
        numbers=none,
        numberstyle=\scriptsize,
        label={#2}, 
        caption={#3},
        escapeinside={(*@}{@*)}
    }
}
{}

\title{Processing Large Datasets of\\Fined Grained Source Code Changes}

\author{\IEEEauthorblockN{Stanislav Levin}
\IEEEauthorblockA{The Blavatnik School of Computer Science\\
 Tel Aviv University\\
Tel-Aviv, Israel\\
stas.levin@cs.tau.ac.il}
\and
\IEEEauthorblockN{Amiram Yehudai}
\IEEEauthorblockA{The Blavatnik School of Computer Science\\
 Tel Aviv University\\
Tel-Aviv, Israel\\
amiramy@tau.ac.il}
}

\date{December 2018}

\begin{document}
\bstctlcite{IEEEexample:BSTcontrol}

\maketitle

\raggedbottom

\thispagestyle{plain}
\pagestyle{plain}

\begin{abstract}


In the era of Big Code, when researchers seek to study an increasingly large number of repositories to support their findings, the data processing stage may require manipulating millions and more of records. 

In this work we focus on studies involving fine-grained AST level source code changes. We present how we extended the CodeDistillery source code mining framework with data manipulation capabilities, aimed to alleviate the processing of large datasets of fine grained source code changes. 
The capabilities we have introduced allow researchers to highly automate their repository mining process and streamline the data acquisition and processing phases.
These capabilities have been successfully used to conduct a number of studies, in the course of which dozens of millions of fine-grained source code changes have been processed.
\end{abstract}

\section{Introduction}

Fine-grained AST level source code changes play an important role in improving our understanding of software maintenance and its evolution \cite{romano2012analyzing,giger2011comparing,fluri2005fine,fluri2009analyzing,marsavina2014studying,levinIcsme2016,levinPromise2017,DBLP:conf/icsm/LevinY17}.
In the era of Big Code, researchers seek to study an increasingly large number of repositories to support their findings. Consequently, their data acquisition stage, where fine-grained source code changes are mined from software repositories, may yield a large dataset of fine-grained source code changes. These datasets can reach millions and more of records. 

In light of the resources (time, memory, etc.) it takes to load, let alone manipulate such large datasets, the task of processing them goes beyond spreadsheets or even the R environment~\cite{R}. In fact, addressing the challenges involved in processing large datasets is one of the driving forces behind many tools and frameworks in the Big Data \cite{diebold2012origin} ecosystem.

In this work we seek to alleviate this task and present the analytical layer we have built and integrated into the CodeDistillery \cite{codeDistilleryRepo} source code mining framework. Given an input dataset of raw fine-grained source code changes produced by CodeDistillery, our analytical layer produces an output dataset consisting of detailed, commit level records. Each record consists of information such as: fine-grained change type frequencies, number of tests suits changed, number of test methods changed, associated task id (if detected) and so on. The output schema is described in detail on CodeDistillery's home page \cite{codeDistilleryRepo}. 

To effectively process large datasets, our analytical layer leverages Apache Spark \cite{sparkSite} (henceforth Spark), a widely popular distributed computation engine. 
The analytical layer we suggest has been successfully used to conduct a number of studies in the field of software maintenance and evolution \cite{levinIcsme2016,levinPromise2017,DBLP:conf/icsm/LevinY17}. This leads us to believe it can be useful for researchers conducting studies that involve fine-grained source code changes. 

\section{Related Work}

Projects such as GH Archive \cite{gharchive}, GH Torrent \cite{Gousi13}, Boa~ \cite{dyer2014mining}, and GitHub API, seek to create scalable services for querying rich source control data. While these tools provide access to commit metadata and even source code, they do not easily support a complex computation such as AST differencing over entire commit histories.

The work that relates most closely to ours, is the experience report by \citet{shang2010experience}, who used MapReduce \cite{dean2008mapreduce} and Pig~\cite{olston2008pig} for scaling software repository mining tools. 
In contrast to \citet{shang2010experience}, we chose to scale our processing using Spark, a modern distributed computation framework.
Our choice is motivated by the advantages Spark offers to its users. In particular, Spark is considered to be more performant~\cite{shi2015clash}, and puts forth a simple, yet powerful programming model~\cite{zaharia2016apache}.

\section{Obtaining fine grained source code changes}

In order to obtain raw fine grained source code changes we use CodeDistillery \cite{codeDistilleryRepo}, a framework built to address the challenges involved in mining fine-grained source code change at scale.
Given a list of Git repository paths (on the filesystem) as an input, CodeDistillery produces an output dataset, consisting of the fine-grained source code changes mined from the specified repositories. This output is generated by replaying the commit history, and recording the AST changes (if any) between consecutive source code revisions (see also the pseudo code in \Cref{lst:mineSingleRepo}). In order to record AST changes, CodeDistillery utilizes the ChangeDistiller AST differencing tool~\cite{fluri2006classifying}.
The output also contains additional metadata
extracted from the commit and the source code files that were processed. 
Using CodeDistillery we were able to gain a 3x performance improvement, compared to sequentially applying ChangeDistiller.

\begin{figure}
\begin{annotatedcsource}{lst:mineSingleRepo}{Mining fine-grained changes from a repository}
mineSingleRepo(repo, outputDir):
  revisions = revisionsOf(repo)
  for (current,next) in pairsOf(revisions)
    changedFiles = fileDiff(current,next)
    for file in changedFiles 
      srcBefore = read(file,current)
      srcAfter = read(file,next)
      astDiff = treeDiff(srcBefore,srcAfter)
      write(astDiff,outputDir)
\end{annotatedcsource}
\end{figure}

\section{Processing large datasets of fine grained source code changes}\label{sec:dataProcessing}

Once a dataset of raw fine-grained source code changes has been acquired, researchers are likely to explore and manipulate it in various ways to pursue insights.
In the following sections we share our experience manipulating large datasets of fine-grained source code changes (acquired using CodeDistillery), and present the capabilities we have built to automate this task.

The analytical layer we have built is implemented in Scala, as a Spark job.
One of the fundamental abstractions in Spark is the resilient distributed dataset (RDD) \cite{zaharia2012resilient}. Spark evaluates resilient distributed datasets lazily, allowing them to find an efficient plan for the user's computation.  
In the code snippets to follow, Spark transformations are highlighted in blue, type annotations are highlighted in violet, and type aliases are in teal. Type annotations for local variables are often omitted in Scala, we explicitly provide them in some of the cases for the sake of clarity.

We begin with reading all the fine-grained source code change datasets which were previously acquired using CodeDistillery, see \Cref{lst:fineGrainedChanges}. 
The way these datasets are organized on disk may vary (e.g., a dataset file per project, a dataset file per a number of projects, etc.), however, the methods we describe hold just the same since the schema of these datasets is consistent. That is, each row (a line in a data file) in any of these datasets has the same format and column structure (as per CodeDistillery's design).
For the sake of demonstration we assume that there is a single dataset file per project, holding all the fine-grained source code changes mined from that software repository.

\begin{figure}
\center
\begin{minipage}{0.95\linewidth}
\begin{annotatedcsource}{lst:fineGrainedChanges}{Reading input data}
type ParsedLine = Array[String]
type Dataset = RDD[ParsedLine]
type MultiDataset = Set[Dataset]

val sparkContext = 
  new SparkContext(
    new SparkConf().setMaster("local[*]"))

val perProjectData: MultiDataset =
(*@\lnote@*)  projects
  (*@\textcolor{black}{.map(prj => }@*)
    sc.textFile(dataOf(prj))
(*@\lnote@*)    .map(line => line.split("#")))

val fineGrainedChanges: Dataset = 
(*@\lnote@*)    sc.union(perProjectData.toSeq)  \end{annotatedcsource}
\end{minipage}
\vspace{1em}
\end{figure}

The variable \verb|projects| (see bookmark~\lnnum{1}~in~\Cref{lst:fineGrainedChanges}) is a collection of project names, over which we iterate and apply a \verb|map| transformation that builds a resilient distributed dataset from each project's data file using Spark's \verb|textFile()| API. 
The data files we read were produced by CodeDistillery, which uses a CSV format with a ``{\small\#}'' (pound) sign as a separator between values. We therefore split each line by the pound sign (see bookmark~\lnnum{2}~in~\Cref{lst:fineGrainedChanges}) to parse the original line into individual values. This parsing results in each line transformed into an \verb|Array[String]|, which we type-alias as \verb|ParsedLine|. Each element in a \verb|ParsedLine| is an individual value parsed from a given line of a given data file. Since the example deals with multiple projects, the \verb|perProjectData| variable ends up being \verb|Set[RDD[Array[String]]]|, which we type-alias as \verb|MultiDataset|.
This \verb|MultiDataset| is then flattened into a \verb|Dataset| using the \verb|union| operation provided by Spark's API (see bookmark \lnnum{3}~in~\Cref{lst:fineGrainedChanges}). 
Each element in the flattened \verb|Dataset| is a \verb|ParsedLine|.
Since resilient distributed datasets are lazy data structures, no actual processing has been done yet. It will only take place once an action (e.g., printing, counting, etc.) is invoked on the \verb|fineGrainedChanges| resilient distributed dataset.

Once the data reading specification is complete, fine-grained source code change frequencies per commit can be computed. That is, how many times each fine-grained source code change type (of the 48 different change types suggested by \citet{fluri2006classifying}, e.g., {\footnotesize ``ADDITIONAL\_CLASS''}, {\footnotesize ``DOC\_INSERT''}), appeared in a given commit (see \Cref{lst:fineGrainedChanges-by-commit}).

\begin{figure}
\begin{minipage}{0.95\linewidth}
\begin{annotatedcsource}{lst:fineGrainedChanges-by-commit}{Computing per commit change type frequencies}
val perCommitFrequencies: 
  RDD[(String, Map[String, Int])] = 
    fineGrainedChanges
    // aggregate per COMMIT_ID
(*@\lnote@*)    .groupBy(vals => vals(COMMIT_ID))
    // compute change types' frequencies                
(*@\lnote@*)    .mapValues(countChangeTypes)
\end{annotatedcsource}
\end{minipage}
\vspace{-1em}
\end{figure}

The per commit change type frequency computation (\Cref{lst:fineGrainedChanges-by-commit}) uses the \verb|groupBy| and \verb|mapValues| transformations.
The \verb|groupBy| transformation takes an element from the RDD it is applied on, i.e., \verb|fineGrainedChanges|, and extracts a key that is used to group all elements with the same key into a single group. Since we would like to compute the frequencies of the different fine-grained source code changes per commit, we specify the key to be the commit id (a.k.a. "commit hash") by passing the \verb|groupBy| a function that extracts the commit id given an element from the RDD (see bookmark~\lnnum{1}~in~\Cref{lst:fineGrainedChanges-by-commit}).
The \verb|groupBy| transformation derives a new RDD where each element is a pair of type \verb|(String, Iterable[Array[String]])|. The first component in this tuple (a.k.a. ``key'') is the commit id, and the second (a.k.a. ``value'') is a collection of all the elements which have this particular key. Then, the \verb|mapValues| transformation is applied (see bookmark~\lnnum{2}~in~\Cref{lst:fineGrainedChanges-by-commit}) on these tuples, which applies a given transformation on each tuple's second component (leaving the tuple's first component unchanged).
The argument passed to \verb|mapValues| (see bookmark~\lnnum{2}~in~\Cref{lst:fineGrainedChanges-by-commit}) is the function to be applied on these tuples' values. It calculates the frequencies of each fine-grained source code change type (see the \verb|countChangeTypes()| function as detailed in \Cref{lst:count-frequency}).

\begin{figure}[H]
\begin{minipage}{0.97\linewidth}
\begin{annotatedcsource}[scala-types-only]{lst:count-frequency}{Computing change type frequencies}
def countChangeTypes(
  parsedLines: Iterable[ParsedLine]): 
  Map[String, Int] =
    parsedLines
(*@\lnote@*)    .map(parsedLine => 
            parsedLine(CHANGE_TYPE))
(*@\lnote@*)    .groupBy(identity)
(*@\lnote@*)    .mapValues(_.size)
\end{annotatedcsource}
\end{minipage}
\end{figure}

The \verb|countChangeTypes| function (see \Cref{lst:count-frequency}) receives an iterable of parsed lines, where each item represents a change performed in a given commit. It returns a a dictionary (\verb|Map[String, Int]|) between the fine-grained source code change type (e.g., {\footnotesize ``ADDITIONAL\_CLASS''}) and its frequency. 
Note that \verb|countChangeTypes| does not operate on RDDs but on Scala's native collections. 

One of the benefits of using Spark's Scala API is that it is consistent with Scala's native collections. In particular, the name and semantics of the \verb|mapValues| and \verb|groupBy| transformations for Scala collections and Spark RDDs are the same.

The \verb|countChangeTypes| function first transforms each parsed line to its corresponding fine-grained source code change type (bookmark~\lnnum{1}~in~\Cref{lst:count-frequency}), and then all the resulting values are grouped using the \verb|identity| key extractor (bookmark~\lnnum{2}~in~\Cref{lst:count-frequency}). This operation forms a dictionary data structure, where the key is a fine-grained source code type and the value is a collection of all its instances.
We then transform these tuples' second component (bookmark~\lnnum{3}~in~\Cref{lst:count-frequency}) by counting the number of instances associated with its key (the change type).
Therefore, \verb|countChangeTypes| returns a dictionary data structure, where the key is a fine-grained source code change type, and the value is this change type's frequency (e.g., {\footnotesize ``ADDITIONAL\_CLASS'' $\rightarrow$ 3}). Consequently, \verb|perCommitFrequencies| (see \Cref{lst:fineGrainedChanges-by-commit}) will hold a resilient distributed dataset of tuples, where the first component is a commit id, and the second component is the returned value of \verb|countChangeTypes|, i.e., the various change types and their frequencies.
The {\ttfamily{perCommitFrequencies}} RDD (see \Cref{lst:fineGrainedChanges-by-commit}) will therefore contain elements of the following form:
\begin{lstlisting}[mathescape=true,numbers=none,frame=none,basicstyle=\ttfamily\small,xleftmargin=0em,backgroundcolor=\color{white},xleftmargin=0em,xrightmargin=0em,framexleftmargin=0em,framexrightmargin=0em,aboveskip=0.5\baselineskip,belowskip=0.5\baselineskip,literate=              {->}{$\rightarrow{}$}{1}]
1a2b3c -> {PARAMETER_INSERT -> 3, DOC_DELETE -> 1}
\end{lstlisting}

\section{Discussion}

In the course of our studies \cite{levinIcsme2016,levinPromise2017,DBLP:conf/icsm/LevinY17}, the data processing stage typically included the following aggregations: commit level; developer level; project level; global statistics.
The analytical layer we present allows researchers to produce commit level aggregations (see \Cref{lst:fineGrainedChanges-aggregate-extension}) and obtain statistics such as: change type frequencies, number of test case (test method) addition/removal/modification, number of test suite (test class) addition/removal/modification, associated ticket id, number of test files, and non test files in a given commit. 

As software analytics \cite{menzies2013software} studies seek to distill large amounts of low-value data into small chunks of very high-value information \cite{menzies2019bad}, we believe that a commit level aggregation strikes the balance for a number of reasons:

\begin{itemize}
    \item The per commit aggregation contains the most information, as some statistics cannot be sensibly computed in the context of per contributor and/or per project aggregations unless transformed. 
    \item The per commit aggregation can be used to derive per contributor, and per project aggregations by performing further grouping.
    \item The per commit aggregation produces a significantly reduced dataset compared to raw fine-grained source code changes. More often than not, it is already sufficiently compact to be further explored and manipulated in interactive environments such as R.
\end{itemize}

We performed a preliminary benchmark where a per commit aggregation was applied on a dataset consisting of fine-grained source code changes acquired from 4 popular open source software repositories: Apache Beam, Apache Hadoop, Apache Camel, and RxJava. This dataset was acquired using CodeDistillery, and consisted of 3,211,933 records with a total size of 2.1 GB.
Our analytical layer was able to complete a commit level aggregation in 73 seconds. In a vanilla R environment (without specialized libraries such as SparkR \cite{sparkR}),
performing a commit level aggregation required 193 seconds, 2.6x slower.
The benchmarks were conducted on macOS Sierra 10.12.6, Intel i7-7820HQ CPU (2.90GHz), and 16 GB RAM (2133 MHz).

\begin{figure}
\begin{minipage}{0.97\linewidth}
\begin{annotatedcsource}{lst:fineGrainedChanges-aggregate-extension}{The new commit level aggregation capability}
// ds1 & ds2 were produced by CodeDistillery
PerCommit.aggregate(Set(ds1, ds2), output)
\end{annotatedcsource}
\end{minipage}
\vspace{-0.8em}
\end{figure}

\section{Conclusion and Future Work}
In this work we presented the analytical layer we built to support the processing of large datasets of fine-grained source code changes. 
We also demonstrated how researchers can obtain commit level aggregations and accumulative statistics from raw datasets produced by the CodeDistillery framework.

Our analytical layer has been successfully used to conduct a number of studies in the field of software evolution and maintenance \cite{levinIcsme2016,levinPromise2017,DBLP:conf/icsm/LevinY17}. 
In the course of these studies it has processed dozens of millions of records.

Future direction may include the exploration of additional analytical layers to be provided on top of the CodeDistillery framework. For example, it could be beneficial to facilitate, or even automate, the building of predictive models using CodeDistillery. Currently, such a task would typically require manual intervention and labour at various points along the way.

\section*{Acknowledgements}
We thank Dr. Boris Levin for his valuable comments and constructive criticism of the manuscript.

\clearpage

\def\UrlFont{\ttfamily\scriptsize}
\def\UrlBreaks{\do\/\do-}
\bibliography{main}
\bibliographystyle{IEEEtranN}

\end{document}